# Advances in Additive Manufacturing of 3D-segmented Plastic Scintillator Detectors for Particle Tracking and Calorimetry

**Contribution to the 25th International Workshop on Neutrinos from Accelerators**

Umut Kose[1], on behalf of 3DET Collaboration

**1** ETH Zurich, Institute for Particle physics and Astrophysics, Zurich, Switzerland
* Umut.Kose@cern.ch

## Abstract

Plastic scintillator detectors with three-dimensional granularity and sub-nanosecond time resolution offer simultaneous particle tracking, identification, and calorimetry. However, scaling to larger volumes and finer segmentation poses significant challenges in manufacturing and assembly due to high costs, extensive time, and precision requirements. To address this, the 3DET R&D collaboration has developed an innovative additive manufacturing approach, allowing for the monolithic fabrication of three-dimensional granular scintillators without the need for additional processing steps. A prototype, featuring a 5 × 5 × 5 matrix of optically isolated scintillating voxels integrated with wavelength shifting fibers, was manufactured and tested using cosmic rays and CERN test beams, demonstrating comparable light yield and reduced crosstalk compared to traditional methods. The developed additive manufacturing technique offers a viable, time-efficient, and cost-effective solution for producing next-generation scintillator detectors, maintaining high performance regardless of size and geometric complexity.

## 1 Introduction

Plastic scintillator detectors with three-dimensional granularity and sub-nanosecond time resolution provide exceptional capabilities for simultaneous particle tracking, identification, and calorimetry. Experiments such as SuperFGD, CALICE, and SOLID develop high-granularity plastic scintillators with larger volumes and finer segmentation to enhance further their detection system. Despite these promising advancements, producing and assembling high-granularity plastic scintillators encounter significant challenges, including prolonged manufacturing times, the necessity for high precision in complex designs, and scaling the production processes. Additive manufacturing emerges as a promising solution to address these challenges by enabling the efficient production of complex detector structures.

The 3D printed Detector (3DET) R&D collaboration was formed to investigate and develop additive manufacturing (AM) techniques capable of producing 3D-printed particle detectors that match the performance of the current state-of-the-art systems. Fused Deposition Modelling (FDM) is considered among various AM methods for its reliability and versatility in producing plastic scintillator particle detectors with diverse and complex geometries.

In the initial phase of the R&D program, efforts focused on validating the concept by defining an optimal material composition. A scintillating filament was formulated with an optimal mixture of polystyrene, pTP, and POPOP, along with plasticizers to introduce flexibility into the scintillating filament. The plasticizer was later removed from the formulation. Test of 3D-printed 1 $cm^3$ samples using cosmic muons and a cesium source demonstrated the light yields comparable to those of traditional cast and extruded scintillators [1]. Moreover, further studies revealed a technical attenuation length of approximately 19 cm in a bar of 5 cm in lengths, which is adequate for fine-granularity scintillator detectors. Additionally, the first 3D-printed matrix of polystyrene cubes, optically isolated with a custom fabricated white reflective filament composed of PMMA mixed with $TiO_2$, produced and showed a low cube-to-cube light crosstalk, less than 2%, and uniform light yield among the cubes. The tolerance of the reflector thickness and cube shape was approximately 0.5 mm [2]. During earlier R&D phases, no studies have attempted to 3D-print hollow structures within cubes to host wavelength shifting (WLS) fibers.

## 2 Advancement in technology: Fused Injection Molding

To address the limitations of traditional FDM techniques in producing plastic scintillator particle detectors, the 3DET collaboration developed the Fused Injection Molding (FIM) technique [3]. As shown in Figure 1, this innovative approach begins with 3D-printing an optically reflective frame using FDM. This frame is pre-configured with voxel shapes and internal cavities to ensure precise geometry and optical isolation between segments, minimizing light crosstalk.

After fabricating the reflective frame, metal rods are inserted through predefined holes to create 1.1 mm in diameter circular voids for the WLS fibers. The cavities are then filled with molten polystyrene (PS) in a controlled bottom-to-top motion, ensuring a uniform filling. A custom liquefaction system is integrated into the FDM setup to achieve a fast, high-quality, and consistent formation of the PS structure. This system features an elongated nozzle to distribute the scintillating material evenly and a spring-pressurized plate to confine the melt pool while allowing air to escape, ensuring void-free cavities.

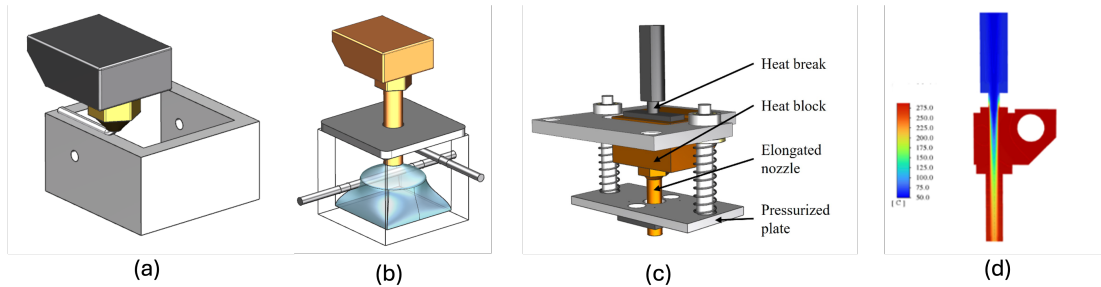

Figure 1: Fused injection modeling: [a] 3D-printing the reflective frame with FDM, [b] Filling voxels with plastic scintillator, [c] Custom design of the extrusion system, [d] CFD simulation showing temperature distributions through melting system.

To optimize the melting process, Computational Fluid Dynamics (CFD) simulations were performed, focusing on melting components, heat block, and nozzle design. The analysis determined optimal operating parameters, including an extrusion speed of 15 mm/s and a heat block temperature of 300 °C, ensuring precise control of molten PS flow and maintaining the material's scintillation properties.

The FIM method allows for the fabrication of complex, precisely defined structures without post-processing, distinguishing it from previous approaches. By eliminating the need for polishing or other subtractive techniques, FIM provides a streamlined, efficient, and cost-effective solution for manufacturing plastic scintillator particle detectors.

### 2.1 3D Printing a Monolithic SuperCube

Using the FIM technique, the 3DET collaboration successfully produced a monolithic SuperCube, a 5 × 5 × 5 matrix of optically isolated scintillating voxels. Each voxel includes precisely defined holes for housing WLS fibers for efficient light transport and collection. The process achieved excellent transparency and optical isolation, as illustrated in Figure 2, which shows the SuperCube under UV light before 3D printing of the top reflective layer.

The reflective frame was constructed using a commercial filament made from white polycarbonate mixed with PTFE, known for its thermal stability up to 300 °C. The frame was designed with horizontal walls of 1.2 mm thickness and vertical walls of 1.5 mm thickness to maintain consistent crosstalk levels across all cubes. This efficient manufacturing process requires no post-processing and allows the detector to be instrumented with readout electronics immediately after the fabrication process is completed.

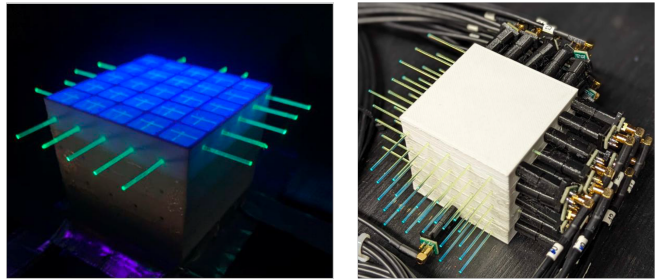

Figure 2: FIM-manufactured SuperCube: [Left] under UV light before 3D-printing top reflective layer and [Right] instrumented with WLS and SiPMs.

## 3  Characterization of Monolithic SuperCube

The FIM-manufactured SuperCube prototype was instrumented with WLS fibers (double-cladding, Kuraray Y11) and silicon photomultipliers (Hamamatsu MPPC 13360-1325CS with 25% photon detection efficiency), and a FERS front-end board managed its readout. The light yield and cube-to-cube crosstalk were measured using cosmic particles and compared to a standard scintillator layer produced via cast polymerization [4].

As shown in Figure 2-a and Figure 2-b, cosmic particles were detected crossing the SuperCube both vertically and diagonally. The 2D projections illustrate the number of photons detected in each readout channel, while the 3D voxels reconstruct the particle tracks. Regarding detection performance, the light yield was comparable to that of scintillators produced by cast polymerization, achieving approximately 28 photoelectrons (p.e.) per channel, with cube-to-cube crosstalk measured at approximately 4%. Further details can be found in Ref. [3].

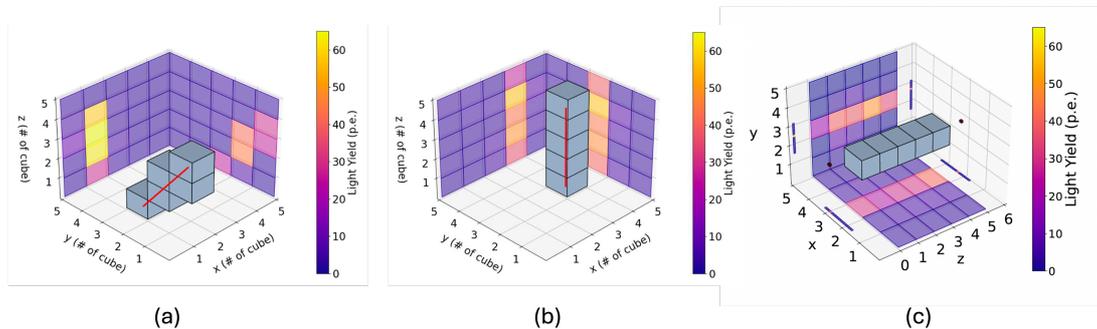

(a)      (b)      (c)

Figure 3: Cosmic particles crossing the SuperCube from top to bottom: [a] diagonal and [b] vertical track. The 2D projections show the number of photons detected in each readout channel. [c] Beam particle crossing central cubes, hits

The SuperCube was subsequently tested at the T9 test beam at the Proton Synchrotron at CERN to validate its performance under beam conditions and to assess the uniformity of its scintillation light yield. A hodoscope, consisting of two X-Y planes of 1 mm$^2$ square scintillating fibers (Kuraray SCSF-78) with a coverage area of 1.6 × 1.6 cm$^2$, provided sub-millimeter resolution for charged particles passing through the central region of the detector to evaluate the uniformity within the cube.

Figure 2-c shows a track of a beam particle detected both in the hodoscopes and the SuperCube prototype. The beam test measured a typical light yield of approximately 28 p.e. per channel, consistent with results obtained from cosmic muons and comparable to samples produced via standard cast polymerization. The cube-to-cube light leakage was observed to be 4–5%, corresponding to 24–30% of the scintillation light escaping through reflective walls. Light yield non-uniformity within a single cube was about 7%, depending on the distance to the WLS fibers, while the variation across the five central cubes was less than 1%, demonstrating excellent stability and reproducibility of the FIM-manufactured scintillator voxels. Further details are discussed in Ref. [5].

## 4 Conclusions

The 3DET R&D collaboration has successfully demonstrated the feasibility of using 3D printing to fabricate plastic scintillator particle detectors with high granularity and integrated holes for WLS fibers in a single monolithic block. This innovative approach eliminates the need for subtractive processes, streamlining production while achieving performance comparable to traditional manufacturing methods such as casting and extrusion.

Ongoing advancements include the development of a heat-resistant, high-performance white reflector to reduce light crosstalk. Efforts are also underway to advance process engineering toward fully automated 3D printing, improving scalability and reducing production costs. The exploration of metal filaments aims to enable the fabrication of the first 3D-printed sampling calorimeter. Additionally, optimization of 3D printing techniques for producing plastic scintillators for neutron capture. These developments highlight the potential of 3D printing technology to revolutionize the production of scintillator-based detectors for a wide range of applications in particle physics and beyond.

## References

[1] 3DET Collaboration, S. Berns et al., A novel polystyrene-based scintillator production process involving additive manufacturing. JINST **15**(10), 10 (2020). doi:10.1088/1748-0221/15/10/P10019.

[2] 3DET Collaboration, S. Berns et al., Additive manufacturing of fine-granularity optically-isolated plastic scintillator elements. JINST bf 17(10), P10045 (2022). doi:10.1088/1748-0221/17/10/P10045.

[3] 3DET Collaboration, T. Weber et al., Additive manufacturing of a 3D-segmented plastic scintillator detector for tracking and calorimetry of elementary particles. https://arxiv.org/abs/2312.04672.

[4] A. Boyarintsev et al., Demonstrating a single-block 3D-segmented plastic-scintillator detector. JINST **16**(12), P12010 (2021). doi:10.1088/1748-0221/16/12/P12010.

[5] 3DET Collaboration, B. Li et al., Beam test results of a fully 3D-printed plastic scintillator particle detector prototype. https://arxiv.org/abs/2412.10174.